\documentstyle[preprint,aps]{revtex}
\begin{document}
\title{Influence of radiative interatomic collisions on
an atom laser}
\author{T.W. Hijmans$^{1}$, G.V. Shlyapnikov$^{1,2}$, and
A.L. Burin$^{2}$}
\address{
{\em (1)} { Van der Waals - Zeeman Institute, University of Amsterdam,\\
Valceniestraat 65-67, 1018 XE Amsterdam, The Netherlands}\\
{\em (2)} { Russian Research Center Kurchatov Institute,\\
Kurchatov Square, 123182 Moscow, Russia}\\
}
\date{\today}
\maketitle
\begin{abstract}
We discuss the role of light absorption by pairs of atoms
(radiative collisions) in the context of a model for an atom laser.
The model is applied
to the case of VSCPT cooling of metastable triplet helium. We
show that, because of radiative collisions,
for positive detuning of the driving light fields
from an atomic resonance the operating conditions for the atom
laser can only be marginally met. It is shown that the system only
behaves as an atom laser if a very efficient sub-Doppler precooling
mechanism is operative. In the case of negative frequency detuning
the requirements on this sub-Doppler mechanism are less restricting,
provided one avoids molecular resonances.
\end{abstract}

\vspace{4mm}

\vspace{4mm}

\section{Introduction}

The investigation of macroscopic quantum phenomena
is one of the prime motivations for the study of
ultra-cold atomic gases. Recent successful experiments on
Bose-Einstein condensation (BEC) in trapped rubidium \cite{Cor95},
lithium \cite{Hul95} and sodium \cite{Ket95} gave a tremendous boost
to this field of research. The breakthrough leading to the achievement
of BEC was
the implementation of evaporative cooling schemes, where
cooling and thermal quasi-equilibrium are provided by interatomic
elastic collisions.
It is important that interaction between atoms also governs
the formation kinetics of a Bose condensate \cite{Kagan92}.

A principal question with regard to quantum
statistical effects concerns the possibility to reach
large or even macroscopic occupation numbers for a single particle
state in a collisionless Bose gas, where cooling and phase-space
compression proceed through interaction of atoms with light rather
than through interatomic collisions.
The central idea is to exploit the bosonic
nature of the particles: once the occupation number of a single
particle state becomes larger than unity, this enhances the rate at
which the state is filled and strongly influences the population
dynamics \cite{Zoller,You}.
This is equivalent to the gain mechanism in lasers.
Accordingly, the term ``atom laser'' \cite{Ols95,Spr95,Wiseman95,Borde95}
has been coined to
describe a gas far from thermal equilibrium, with atoms
accumulating in a single quantum state or at least in a very
small region of phase space.

Optical cooling of a gas in the collisionless regime requires the
condition
\begin{equation}      \label{n}
n\lambdabar^3\ll 1,
\end{equation}
where $n$ is the gas density and $2\pi\lambdabar$ the wavelength of light.
In the opposite limiting case the evolution of excited atomic states
is mainly governed by interatomic collisions induced by resonance
dipole interaction \cite{Vdovin}. These collisions proceed at a rate
much larger
than the rate of spontaneous emission and destroy cooling.
In view of Eq.~(\ref{n}), to achieve occupation
numbers of the order of unity or larger the atoms should gather
in a momentum range smaller than the single photon recoil. Cooling
schemes based on dark states, such as
Velocity Selective Coherent Population Trapping (VSCPT)
\cite{Asp88,Law94,Law95} or
schemes involving Raman pumping between different hyperfine states
\cite{Kas92} can be used for this purpose.
Whatever the scheme chosen, it will necessarily involve
laser fields which drive the pumping, and hence spontaneous emission
which, under certain conditions, can lead to compression of atoms in
momentum space. Reabsorption of spontaneously emitted photons,
destroying the momentum-space compression, can be, hopefully,
circumvented by selecting a high frequency detuning of driving light
\cite{Ols95} or by choosing at least one of the sample dimensions smaller
than the mean free path of a photon. Such geometrical means of
reducing the effect of multiple reabsorption were successfully exploited
in the optical cooling of atomic hydrogen \cite{Set93} and have been
discussed in the context of atom lasers \cite{Dal95}.

Exchange of longitudinal virtual photons between excited and
ground-state atoms leads to resonance dipole interaction.
The influence of the mean field of this interaction on VSCPT
cooling was found small under the condition Eq.~(\ref{n})
\cite{Wallis}.
We will consider the case where the frequency detuning
$\delta$ of the laser fields from resonance with an excited
state is large compared to the natural line width $\Gamma$ of
this state.
In this case resonance dipole interaction manifests itself in
interatomic pair collisions (see e.g., \cite{Kag88}).
At interparticle distances where resonance dipole interaction
compensates the frequency detuning a colliding pair is resonant
with the light. This is the origin of the well known process of
light absorption in pair collisions, which we will refer to as radiative
collisions.
Of particular interest is the limit of ultra-cold collisions,
reviewed in e.g. ref. \cite{Julienne}.

At ultra-low temperatures corresponding to the $s$-wave scattering
limit, under the condition of Eq.~(\ref{n}) the rate of radiative
collisions is normally much smaller than the rate of absorption of light
by single atoms.
The situation changes drastically for atoms in so called dark
states which are decoupled from the driving light field.
In this case the coupling of the dark-state atoms with the driving light
is induced by their radiative collisions with the atoms in coupled
states. This process, unlike reabsorption of spontaneously emitted
photons, can not be eliminated by arranging the sample geometry.
In this paper we argue that for $|\delta|\gg \Gamma$
at any gas density the phase-space compression to occupation
numbers larger than unity can be frustrated by this type of
radiative collisions.
The physical
reason that these collisions can be important even in a very dilute gas
is the following:
If we consider a small sphere of
radius $\tilde p$ around $p=0$ in momentum space, the filling rate of this
momentum region from the gas cloud of spatial density $n$ is
roughly proportional to $(1+n(\tilde p))\tilde p^3n$, where
$n(\tilde p)$ is the characteristic occupation number of momentum
states in the small sphere.
The collisional loss of atoms from this sphere is proportional
to $n(\tilde p)\tilde p^3 n$.
The ratio of filling to loss rate, being large for small occupation
numbers, decreases with increasing $n(\tilde p)$ and tends to a constant
for $n(\tilde p)>1$. The key point is that both the filling rate and
the rate of radiative collisions depend on the degree of excitation of
the atoms in a similar way. However, the phase space available for the
filling is ultimately set by the momentum change in the last
spontaneous emission event, i.e. the photon momentum, whereas the
phase space available for collisions can be much larger. In view of
this the question whether radiative collisions can prevent the
occupation numbers from reaching unity is principal.

We will analyze a simple general model for an atom laser which
includes radiative collisions and apply the results to
VSCPT cooling on a J=1 to J=1 transition, successfully used to cool
helium atoms in
the metastable $2^3S_1$ state to below the recoil energy
\cite{Asp88,Law94,Law95}.
For this scheme we discuss both the case of large positive and large
negative detuning and show that the latter case is more
promising for realizing an atom laser.

\section{General scheme}

We consider a general scheme for an atom laser, such as presented in
ref. \cite{Ols95}. This scheme is depicted in Fig.\ \ref{1}.
All atoms are confined to
a volume much larger than the optical wavelength and such that
the eigenstates of the translational motion of the
atoms are (approximate) momentum eigenstates.
We have two sets of atomic states called the system (labeled $s$) and
bath (labeled $b$).
The system $s$ comprises just the atoms in the compressed momentum
space described above.
The laser fields are tuned close to a resonance
involving an excited atomic state $e$, which optically pumps the atoms
from the bath into the system. In Fig.\ \ref{1} the pumping
scheme is schematically represented as a single transition connecting
the bath states to the excited state followed by spontaneous emission
into the system states, although in practice the pumping scheme can be
more complicated.
Spontaneous emission makes the atom end
up in a system state with momentum ${\bf p}$, chosen so to say "by
chance" as
a result of the random direction of the emitted photon.
As the focus of this paper is the role of radiative collisions with
bath atoms, we
omit all other potential loss mechanisms
(in contrast to e.g. ref. \cite{Ols95}) such as
decay of the system
states due to absorption of spontaneously emitted photons.
Then the rate equation for the occupation numbers $n_s({\bf p})$ of
the system states with momentum ${\bf p}$ takes the form:
\begin{equation}
\!\dot{n}_s({\bf p})\!=\!-\Gamma_s({\bf p}) n_s({\bf p})\!+
\!(1+n_s({\bf p}))
\int \frac{d {\bf q}}{8 \pi^3}n_b({\bf q})\Gamma_b({\bf q})
P({\bf p},{\bf q})
-n_s({\bf p})
\int \frac{d {\bf q}}{8 \pi^3} G({\bf p},{\bf q}) n_b({\bf q}).
\label{evolution}
\end{equation}
The first term denotes the loss rate of
isolated atoms in the system due to the presence of the light fields.
It involves optical pumping from the system states back into the bath.
Clearly it is advantageous if the lifetime of systems states increases
with decreasing momentum. We will assume that this repumping rate
vanishes quadratically with
momentum for small $p$: $\Gamma_s({\bf p})=\tilde\Gamma \ (p/p_*)^2$. Such a
quadratic dependence is a naturally encountered in schemes based on
VSCPT or on velocity selective Raman pumping. The region
$p<p_*$ can be called a trap in momentum space. The second term in
Eq.~(\ref{evolution}) is the pumping rate from bath states into the
system. Here $\Gamma_b({\bf q})$ is the probability per unit time
that an atom leaves the bath state having momentum ${\bf q}$,
$n_b({\bf q})$ is the occupation number of the
corresponding bath state, and $P({\bf p},{\bf q})$ is the probability
density for a bath atom with momentum ${\bf q}$ to end up in a system
state with momentum ${\bf p}$ after a spontaneous emission event.
The prefactor $1+n_s({\bf p})$ in the second term is the
Bose enhancement factor which is responsible for the "lasing" process.
The third term in Eq.~(\ref{evolution}) is the focus of this paper.
It describes the absorption and subsequent reemission
of a photon from the laser fields in a
collision between the system atom and the bath atom.
As the photon absorption strongly changes the relative motion of
colliding atoms, such radiative collisions will be a loss mechanism
for system atoms trapped in the space of low momenta $p < p_*$.
For large frequency detuning $\delta$ the rate constant of radiative
collisions $G({\bf p},{\bf q})$ becomes momentum independent and can
be written as $G=\beta\tilde\Gamma\lambdabar^3$. The cubic dependence on
$\lambdabar$, and the proportionality to $\tilde\Gamma$ are a consequence of
the resonant dipole interaction. The coefficient
$\beta$ depends on $\delta$, the Rabi frequency and
details of the level structure.

In view of Eq.~(\ref{n}) it is natural to assume that the momentum trap
size $p_{*}<k$, where $k$ is the photon momentum.
In order to simplify the picture we describe the bath by introducing
a sphere $p_{max}$ in momentum space,
with the bath occupation numbers independent of ${\bf p}$ for
$p<p_{max}$ and zero otherwise.
Also $\Gamma_b$ is momentum independent.
Accordingly, $n_b({\bf q})\equiv n_b \Lambda_b^3$, where $n_b$ is the
real space density of bath atoms, and their De Broglie wavelength
$\Lambda_b=(6 \pi^2)^{1/3}/p_{max}$.
In a non-thermal gas, the kinetic energy of bath atoms is
maintained by a dissipative optical cooling mechanism and
thus higher than the recoil energy, i.e., $p_{max}>k$.
Since the momentum change during spontaneous
emission is of
order ${\bf k}$ and $p_*<k$, the integral in the second
term of Eq.~(\ref{evolution}) can be written as
$\alpha\Gamma_b n_b\Lambda_b^3$, where $\alpha$ is a numerical
coefficient which depends on the cooling scheme and the level structure
of the atoms.
With $\Gamma_b\approx \tilde\Gamma$, the rate equation (\ref{evolution})
now reduces to:
\begin{equation}
\dot{n}_s({\bf p})\!=\!\tilde\Gamma [-(p/p_*)^2 n_s({\bf
p})\!+\!\alpha  n_b \Lambda_b^3 (1\!+\!n_s({\bf p}))\!-
\!\beta n_s({\bf p})n_b \lambdabar^3].
\label{simpelevol}
\end{equation}
We introduce the parameter $\eta\equiv
(\alpha  \Lambda_b^3)/(\beta \lambdabar^3)$.
Apart from the ratio $\alpha/\beta$, the parameter $\eta$ essentially
denotes the ratio
of the recoil energy $E_r=k^2/2m$ ($\hbar$ is set equal to unity
throughout) to
the bath ``temperature'' $T_b\sim 2\pi/ m\Lambda_b^2$. As we necessarily
have $T>E_r$ ($\Lambda_b<\lambdabar$) we may
expect $\eta$ to be less than unity.
For $\eta<1$ we find a steady state solution for the occupation
numbers in the trap:
\begin{equation}             \label{steady state}
n_s({\bf p})=\eta/[ (p/p_*)^2/\beta n_b \lambdabar^3 + (1-\eta)].
\end{equation}
The maximum occupation number is achieved for $p\rightarrow 0$.
It is smaller than
unity unless $\eta$ is very close to 1. The fraction of
particles accumulated in the momentum-space trap is of the order of
$(p_*/p_{max})^3\ll 1$, hence for
$\eta<1$ the atoms predominantly remain in bath states.

In the case $\eta>1$ there is no steady state solution.
The occupation numbers of states with
$(p/p_*)^2<(\eta-1) \beta n_b \lambdabar^3$  grow
exponentially,
with a characteristic inverse growth time $\tilde\Gamma (\eta-1) \beta n_b
\lambdabar^3- \tilde\Gamma (p/p_*)^2$.
We have an atom laser.
There is no threshold due to the fact that we omitted all
loss mechanisms except radiative collisions.
Ultimately the bath will be depleted and the above approximations
break down.
Clearly, two criteria have to be
met in order to make the atom laser work: in order to
have $\eta>1$ the prefactor $\beta$ which governs radiative
collisions should be much smaller than $\alpha$ and the bath
temperature should be kept as close to $E_r$ as possible.

\section{Application to He$^{*}$}

We will apply the general scheme for the atom laser
described above to the case of VSCPT
cooing of He in the metastable $2^3S$ state (He$^{*}$)
In Fig.\ \ref{2} we show the relevant levels involved in VSCPT cooling
of He$^{*}$. For simplicity we will first consider a one-dimensional
picture and later generalize it to 3-d. The model in the previous
section relates only to the 3-d case. The 1-d
calculation presented here is not meant as a 1-d variant of this model
but serves only to obtain numerical
results which we will show to be independent
of the dimension and which we will subsequently use in the generalization to
3-d.

In the 1-d VSCPT case the sample is irradiated with a $\sigma_+$
and a $\sigma_-$
polarized beam propagating in the positive and negative $z$
direction, respectively.
The Hamiltonian of interaction of an isolated atom with the light
field has $6$ eigenstates.
Optical
pumping ensures that after a comparatively short time only three of
the states in Fig.\ \ref{2} remain populated \cite{Asp88}.
In the absence of light the wavefunctions of two of these states can be
written in the form:
\begin{equation}          \label{singlestates}
\!\chi_{c,u}({\bf p})\!=\!\frac{1}{\sqrt{2}}[\chi_1\!\exp\{i({\bf p}\!
+\!{\bf k}){\bf R}\} \pm
\chi_{-1}\!\exp\{i({\bf p}-{\bf k}){\bf R}\}].
\end{equation}
Here the labels $c$ and $u$ stand for coupled and uncoupled states,
the plus sign relating to $\chi_c({\bf p})$
(using the phase convention of ref. \cite{Asp88}).
The atom coordinate and momentum are ${\bf R}$ and ${\bf p}$,
and $\chi_M$ is the wavefunction of
the $2^3S$ atomic state with spin projection $M$ on the direction of
light propagation, which we select as quantization axis. The state
$\chi_0$ in Fig.\ \ref{2} is depopulated by optical pumping. The states
$\chi_1$ and $\chi_{-1}$ are coupled by the $\sigma_-$ and $\sigma_+$
beams, respectively, to the excited $2^3P_1$ state $\phi_0$, with
zero projection of
the total electron angular momentum. The wavefunction of this
state can be written as $\phi_0({\bf p})=\phi_0\exp(i{\bf pR})$.

The state $\chi_u({\bf p})$ is called uncoupled because in the limit of
$p\rightarrow 0$ it is completely decoupled from the driving light
fields.
The rates $\Gamma_c$ and $\Gamma_u$ at which $\chi_c({\bf p})$ and
$\chi_u({\bf p})$
scatter photons are given by \cite{Asp88}:
\begin{eqnarray}
&\Gamma_c&=(\Omega^2/2)\frac{\Gamma}{\delta^2+\Gamma^2/4},
\label{Gamma}\\
&\Gamma_u&(p)=(k p/m)^2\frac{\Gamma}{2 \Omega^2}; \hspace{1cm} p \alt p_*,
\nonumber \\
&\Gamma_u&(p)=\Gamma_c; \hspace{2cm} p \agt p_*,
\label{Gamma_-}
\end{eqnarray}
where $p_*=\Omega^2 m / k \delta$.
The Rabi frequency is defined as
$\Omega=dE$, where $d$ is
the dipole moment of the $2^3S-2^3P_1$ atomic transition and
$E$ the electric field amplitude for each beam.
The scattering rate for atoms in the state $\chi_u({\bf p})$ is
proportional to $p^2$ for small momenta.
Hence we can identify the states
$\chi_u({\bf p})$ for $p<p_*$ as our system states.
The states $\chi_c({\bf p})$, as well as the states $\chi_u({\bf p})$
with $p>p_*$, can be
considered as comprising the bath.
To complete the correspondence with
Eqs.~(\ref{evolution}) and (\ref{simpelevol}) we note that the excited state
decays into $\chi_{1}$ and $\chi_{-1}$ (and hence into $\chi_c$ and
$\chi_u$)
with equal probability.
Accordingly, the coefficient $\alpha=1/2$
(see also ref. \cite{Asp88}).

Light absorption in pair interatomic collisions requires at least one
of the colliding atoms to be in a coupled state, since for a pair of
atoms both in
uncoupled states the resonance dipole interaction is practically
absent.
Therefore, the collisional loss term for the system atoms in
Eq.~(\ref{evolution}) will be proportional to the occupation number
$n_c(k)$ of the coupled states with momenta around $k$.
We defined our bath as containing
both coupled and uncoupled states, but only the coupled part contributes
to the radiative collisions. Except for very small momenta,
the time scale on which
the populations change is long compared to the optical pumping
time, therefore
detailed balance ensures that the ratio of the occupation numbers of
coupled and uncoupled states satisfies the condition
\begin{equation}                    \label{detailed}
n_c(p)\Gamma_c(p)=n_u(p)\Gamma_u(p).
\end{equation}
Hence, as $\Gamma_c=\Gamma_u=\tilde\Gamma$ for $p>p_*$, we
can express the decay of $n_s(p)$ in terms of the bath
occupation numbers by substituting $n_c(p)=n_b(p)/2$.

The effect of radiative collisions can
not be reduced by relaxing the above assumption $p_*<k$. Let us
demonstrate this for the extreme case, where $p_*\agt p_{max}$. We
still assume $\Omega\ll \delta$.
In this case it is more natural to define the system states as
uncoupled states in a small momentum range
$p<\tilde p$ near zero ($\tilde p<k$). Then, for bath states we
have $\Gamma_u(q)\approx\Gamma_c(q/p_*)^2$ and, hence, most of the
atoms will be pumped into the uncoupled state (see Eq.~(\ref{detailed})).
As only the population of the coupled
states contributes to the collisional loss of
system atoms, the rate of radiative collisions involving coupled-state
atoms with momentum $q$ is reduced by a factor
$\sim (q/p_*)^2$ compared to the case $p_*<k$. On the other hand, the
optical pumping rate is reduced by the same factor.
This is again clear from Eq.~(\ref{detailed}) which shows
that the optical pumping rate from coupled states should be
exactly half the total pumping rate from coupled and uncoupled
states, just as we found above for $p_*<k$.

\section{Radiative collisions. Positive detuning}

Let us now consider light absorption in pair collisions of atoms in
the uncoupled state $\chi_u({\bf p})$ with atoms in the coupled state
$\chi_c({\bf p'})$.
We will first discuss the case of positive frequency detuning $\delta$,
where the light is at resonance with continuum states of the
excited quasimolecule.
The Hamiltonian of resonance dipole interaction for a pair of atoms
labeled by $(1)$ and $(2)$ is given by
\begin{equation}            \label{hamdip}
\hat V =\frac{(\hat{\bf d}^{(1)}\hat{\bf d}^{(2)})r^2-
3(\hat{\bf d}^{(1)}{\bf r})(\hat{\bf d}^{(2)}{\bf r})}{r^5},
\end{equation}
where $\hat {\bf d}^{(1)}$ and $\hat {\bf d}^{(2)}$ are the dipole
moment operators of the colliding atoms, and ${\bf r}$ the vector of
interparticle separation.
Under the condition $\delta \gg \Gamma=4d^2/3\lambdabar^3$ radiative
transitions predominantly occur at interparticle distances
$r\ll\lambdabar$ where the resonance dipole interaction $V\propto d^2/r^3$
compensates the frequency detuning. At such distances we can omit the
factors $\exp[i({\bf p}+{\bf k}){\bf R}]$ and $\exp[i ({\bf p}-{\bf
k}){\bf R}]$
in the expressions for $\chi_{c,u}({\bf p})$, and the initial-state wavefunction
of a colliding pair takes the form
\begin{equation}             \label{initial}
\Psi_i=\hat P_g\chi_u^{(1)}\chi_c^{(2)},
\end{equation}
where $\hat P_g$ is the symmetrization
operator with respect to interchange of electrons and their inversion.
The index $g$ shows that the initial electronic state of the quasimolecule
is {\it gerade}.
The two atoms forming the pair are labeled by the superscripts (1) and
(2).

Excited quasimolecular states to which  radiative transitions occur are
{\it ungerade}. For $\delta>0$ the quasimolecule formed in the
light absorption process corresponds to repulsive potential of
interaction.  Diagonalizing the Hamiltonian of resonance dipole
interaction Eq.~(\ref{hamdip}) we find five such states:
\begin{eqnarray}                 \label{states}
\tilde\Phi_{21} & = & \hat P_u\frac{1}{\sqrt2}(\tilde\chi_1^{(1)}
\tilde\phi_0^{(2)}+
\tilde\chi_0^{(1)}\tilde\phi_1^{(2)}),  \nonumber\\
\tilde\Phi_{2-1} & = & \hat P_u\frac{1}{\sqrt2}(\tilde\chi_{-1}^{(1)}
\tilde\phi_0^{(2)}+
\tilde\chi_0^{(1)}\tilde\phi_{-1}^{(2)}),  \nonumber\\
\tilde\Phi_{11} & = & \hat P_u\frac{1}{\sqrt2}
(\tilde\chi_1^{(1)}\tilde\phi_0^{(2)}-
\tilde\chi_0^{(1)}\tilde\phi_1^{(2)}),  \nonumber\\
\tilde\Phi_{1-1} & = & \hat P_u\frac{1}{\sqrt2}
(\tilde\chi_{-1}^{(1)}\tilde\phi_0^{(2)}-
\tilde\chi_0^{(1)}\tilde\phi_{-1}^{(2)}),  \nonumber\\
\tilde\Phi & = & \frac{1}{\sqrt{6-2\sqrt3}}
\left( \sqrt{2}\tilde\Phi_{20}+
(\sqrt{3}-1)\tilde\Phi_{00}\right)  \nonumber\\
&=& \hat P_u\frac{1}{\sqrt{6-2\sqrt3}}
(\tilde\chi_1^{(1)}\tilde\phi_{-1}^{(2)}+
(\sqrt 3 -1)\tilde\chi_0^{(1)}\tilde\phi_0^{(2)}+
\tilde\chi_{-1}^{(1)}\tilde\phi_1^{(2)}),
\end{eqnarray}
where $\tilde\Phi_{JM}$ is the electron wavefunction of the excited
($2^3P_1-2^3S_1$)
quasimolecular state, with total electron angular momentum $J$ and projection
$M$, and $\tilde\phi_m$ is the wavefunction of the $2^3P_1$
atom, with projection $m$ of the total angular momentum.
The tilde is used to denote that the quantization axis is here the
internuclear axis.
The first four states are characterized by the potential
\begin{equation}             \label{pot1}
V(r)=\frac{d^2}{2r^3},
\end{equation}
and the fifth one by
\begin{equation}             \label{pot2}
V_{*}(r)=\left( \frac{\sqrt3 +1}{2}\right) \frac{d^2}{r^3}.
\end{equation}
One can transform the states of Eq.~(\ref{states}) on the original
quantization axis
(direction of light propagation) by
using the transformation
\begin{equation}               \label{trans}
\tilde\Phi_{JM'}=\sum_{M}(-1)^{M'-M}
\Phi_{JM}D^{J}_{-M',-M}(\theta,\varphi,0),
\end{equation}
where $D^J_{M'M}(\theta,\varphi,\phi)$
is a finite rotation matrix. The angles $\theta$, $\varphi$
determine the orientation of the internuclear axis with respect to the axis
of quantization.

Radiative transitions couple the initial state with
the states  $\Phi_{11},
\Phi_{1-1},\Phi_{21},\Phi_{2-1}$. The dipole moment
of the corresponding transitions is equal to $d/\sqrt2$.
Accordingly, the dipole
moment $d_{JM}$ of transitions from the initial state (\ref{initial}) to the
first four states (\ref{states}) is
\begin{equation}                  \label{dip}
d_{JM}=\frac{d}{\sqrt2}(D^J_{-M,-1}+D^J_{-M,1}).
\end{equation}
The dipole moment of the transition to the state $\tilde\Phi$ is
\begin{equation}                   \label{dipp}
d_{*}=-\frac{d}{\sqrt{6-2\sqrt3}}(D^2_{0,-1}+D^2_{01}).
\end{equation}
In our limit of large detuning the light absorption is
dominated by distances in a narrow vicinity of the resonance separation
$r_{\delta}$ determined by
the condition $V(r)=\delta$ (or $V_{*}(r)=\delta$), and the
number of absorption events per unit time and unit volume is
given by (see e.g. \cite{Kag88}):
\begin{equation}            \label{rate1}
\nu=2\pi\Omega^2 n_c n_s\int d^3r\left( \left| \frac{
d_{*}(\theta,\varphi)}{\sqrt{2}d}\right|^2
\delta(V_{*}(r)-\delta)+\sum_{JM}\left|
\frac{d_{JM}(\theta,\varphi)}{\sqrt{2}d}\right|^2
\delta(V(r)-\delta)\right) ,
\end{equation}
where $n_s$ is the density of atoms in system states (uncoupled states
with $p<p_{*}$), $n_c$ is the density of coupled states,
and the summation should be
performed over the first four states (\ref{states}).
Using Eqs.~(\ref{dip}), (\ref{dipp}), (\ref{pot1}) and (\ref{pot2})
we obtain from Eq.~(\ref{rate1}):
\begin{equation}               \label{rate2}
\nu=7.4\left( \frac{\Omega}{\delta}\right)^2\Gamma n_s
(n_c\lambdabar^3).
\end{equation}

If we replace $n_s$ in Eq.~(\ref{rate2}) by $n_s(\bf{p})$ we obtain
the decay rate of the occupation number of system atoms with
momentum ${\bf p}$ due to pair radiative collisions with bath atoms in
uncoupled states.
As those represent only a part of the bath, the above defined effective
rate constant of radiative collisions $G$ (and parameter $\beta$) are
proportional to the ratio $n_c/n_b$. As we already mentioned
for $p_*<k$ this ratio is equal to $1/2$.
Then, comparing Eqs.~(\ref{rate2}), (\ref{Gamma}) and (\ref{Gamma_-}) with
Eq.~(\ref{simpelevol}) we find $\beta=7.4$ and (with
$\alpha=1/2$) obtain
\begin{equation}               \label{eta}
\eta=0.068(\Lambda_b^3/\lambdabar^3)=4.0 k^3/p_{max}^3\approx 3
(E_r/T_b)^{3/2}.
\end{equation}
Unless $p_{max}$ is within a factor $1.5$  of $k$,
the parameter $\eta$ is less
than unity. In other words the atom laser can only be realized in the
case of positive $\delta$ if the bath is essentially cooled down to the
recoil energy. As noted above Eq.~(\ref{eta}) remains unchanged for $p_{*}>k$
since in this case the ratio $n_c/n_b$ becomes smaller
than $1/2$, but the filling rate reduces by the same factor.

The above results are easilly generalized to 3-d and perhaps
surprisingly the result does not change. The key point is that the
final states of the colliding pair given in Eq.~(\ref{states})
remain unchanged and we need only to reconsider the initial state.
In a 3-d VSCPT cooling scheme different
configurations of laser fields are possible. It has been shown
\cite{Ols92} (see also ref.\cite{Law94}) that
for a $J=1$ to $J=1$ transition there always exists an uncoupled
state with ${\bf p}=0$. This uncoupled state is a vector
\mbox{\boldmath $\psi$}$({\bf R})$ which
satisfies the condition that the
local spin vector is everywhere proportional and parallel to the
polarization vector of the applied light field:
\begin{equation}
\mbox{\boldmath $\psi$} ({\bf R}) = c {\bf E}({\bf R}), \label{psi}
\end{equation}
where $c$ is a normalization coefficient, and ${\bf E}({\bf R})$ is
the laser electric field at position ${\bf R}$.
Commonly the geometry of the light fields is selected such that it
consists of three, mutually orthogonal, pairs of light fields each
consisting of counterpropagating $\sigma_+$ and $\sigma_-$ beams, just
as in the 1-d case. The resulting field ${\bf E}({\bf R})$ is rather
complicated, giving rise to a light field potential and a pumping rate
which are modulated in real space. As in the 1-d case, one can
generalize Eq.~(\ref{psi}) and obtain the expression for the
``uncoupled state'' with ${\bf p} \neq 0$. Then the loss rate from such
"uncoupled" states is again proportional
to $p^2$ \cite{Ols92,Law94}.
When considering radiative collisions we can again omit all
momentum labels, since for $|\delta|\gg \Gamma$ the dominant contribution
to the rate of light absorption comes from
interatomic distances $r\ll \lambda$, while the optical potential (and
the function \mbox{\boldmath $\psi$}) vary
on a length scale of order $\lambda$.
At each point
${\bf R}$ we can define a local quantization axis perpendicular to the
vector \mbox{\boldmath $\psi$}$({\bf R})$ and find two orthogonal
coupled states $\chi_{c1}$ and
$\chi_{c2}$ which form the
complement of \mbox{\boldmath $\psi$}$({\bf R})$.
The uncoupled state plays the role of the state $\chi_u$, introduced
above for the 1-d case (see Eq.~(\ref{singlestates})), and the two coupled
states correspond to superpositions of $\chi_c$ and $\chi_0$.

One difference from the true 1-d case is that due to the
non-local nature of the pumping process neither of the states
$\chi_{c1}$ and $\chi_{c2}$ is depopulated. However, as $\chi_{c1}$
and $\chi_{c2}$ are related by a simple unitary transformation, one
immediately finds that the probability that the atom is optically
pumped into the state \mbox{\boldmath $\psi$}$({\bf R})$ from either
of the coupled states is exactly $1/2$, just as in the 1-d case
described above. Similarly, the
result of the calculation presented above for the rate of radiative
collisions caries over without change:
We should only replace $n_c$ in Eq.~(\ref{rate2}) by the total
density of atoms in the states $\chi_{c1}$ and $\chi_{c2}$.

The results of this section rely on perturbation theory and do not
take into account the influence of light on the wavefunction of the
relative motion of atoms in the initial state. In fact this
wavefunction was implicitly put equal to unity at $r$ close to the
resonance separation $r_{\delta}$, which assumes that the ratio
$\Omega/\delta$ is sufficiently small.
The situation is different if $(\Omega/\delta)kr_{\delta}\gg 1$,
where $k=\sqrt{m\delta}$ is the momentum of the relative motion
acquired by a colliding pair in the light absorption process.
Then the light will provides a repulsion between the potential curve
$V(r)$ (or $V_{*}(r)$) and the potential curve of
the ground electronic state (shifted by the photon energy).
This decreases the probability for two atoms to approach each other to
distances $r\sim r_{\delta}$ where the light absorption is most efficient.
Hence, in principle, there is a possibility to reduce the rate of radiative
collisions by increasing $\Omega/\delta$.
In the case of He$^{*}$ for realistic frequency detuning this requires
the Rabi frequency to be significantly larger than $\delta$, as the quantity
$kr_{\delta}$ will not be much greater than unity.

\section{Negative detuning}

For negative values of $\delta$ the situation is completely
different from the
case described above, as the light can only be at resonance with
discrete vibrational levels (having orbital angular momentum equal to
$1$ and high vibrational quantum number) of the electronically excited
molecule.
This means that radiative collisions will be nothing else than
photoassociation, a process well investigated in ultra-cold alkali
atom gases (for a review see \cite{Heinzen,Lett}).
In analogy to Eq.~(\ref{states}), diagonalizing the Hamiltonian of
resonance dipole interaction Eq.~(\ref{hamdip}) we find four attractive
excited electronic states:
\begin{eqnarray}             \label{atstates}
\tilde\Phi_{22} & = & \hat P_u \tilde\chi_1^{(1)}\tilde\phi_1^{(2)},  \nonumber\\
\tilde\Phi_{2-2} & = & \hat P_u \tilde\chi_{-1}^{(1)}\tilde\phi_{-1}^{(2)}, \nonumber\\
\tilde\Phi_{10} & = & \hat P_u\frac{1}{\sqrt2}(\tilde\chi_1^{(1)}\tilde\phi_{-1}^{(2)}-
\tilde\chi_{-1}^{(1)}\tilde\phi_1^{(2)}),  \nonumber\\
\tilde\Phi_- & = & \frac{1}{\sqrt{6+2\sqrt3}}\left(
(\sqrt{3}+1)\tilde\Phi_{00}-\sqrt{2}\tilde\Phi_{20}\right)  \nonumber\\
&=& \hat P_u\frac{1}{\sqrt{6+2\sqrt3}}
(\tilde\chi_1^{(1)}\tilde\phi_{-1}^{(2)}-(\sqrt 3 +1)\tilde\chi_0^{(1)}\tilde\phi_0^{(2)}+
\tilde\chi_{-1}^{(1)}\tilde\phi_1^{(2)}).
\end{eqnarray}
The first three states are characterized by the
interaction potential $V_{-}(r)=-d^2/r^3$ and the fourth
one by $V_{*-}(r)=(1-\sqrt{3}) d^2/ 2 r^3$.
The transition dipole moments $d_{JM}$
to the first three states of Eq.~(\ref{atstates}) are again given by
Eq.~(\ref{dip}) and the dipole moment of the transition to the state
$\tilde\Phi_-$ is
\begin{equation}                   \label{dipp-}
d_{*-}=\frac{d}{\sqrt{6+2\sqrt3}}(D^2_{0,-1}+D^2_{01}).
\end{equation}
The exact location of discrete vibrational levels
in these potentials can only be found if one knows the short-range
form of the interaction potentials.
Nevertheless, the spacing $\Delta\varepsilon_{\nu}$ between adjacent
levels with
binding energies $\varepsilon_{\nu}$ and $\varepsilon_{\nu+1}$
is determined by the above given resonance dipole potentials $V_{-}$
and $V_{*-}$: $\Delta\varepsilon_{\nu}\sim \varepsilon_{\nu}(r_t/r_0)^{1/2}
\sim (\varepsilon_{\nu}/\Gamma)^{5/6}(\lambdabar/r_0)^{1/2}$,
where $r_t$ is the outer turning point for the relative motion of atoms
in the bound state with vibrational quantum number ${\nu}$, and
$r_0=md^2\gg r_t$.
Hence, we can find the photoassociation rate as a function of
the frequency detuning from the nearest vibrational resonance.
If the light is nearly resonant with the vibrational level $\nu$,
the rate of photoassociation will be
\begin{equation}
\nu_{pa}=4\pi a_J\Omega^2
\frac{\Gamma}{(\delta-\varepsilon_\nu)^2+\Gamma^2}
\left| \int_0^{\infty}\psi_{\nu}(r)\psi_g(r)r^2dr
\right|^2 n_c n_s.  \label{zeronegative}
\end{equation}
Here $\psi_{\nu}(r)$ and $\psi_g(r)$ are the radial wavefunvctions of the
relative motion of atoms in the initial and final electronic states of
the quasimolecule. The coefficient $a_J=1/(2J+1)$ for the states
$\tilde\Phi_{JM}$ given in Eq.~(\ref{atstates}) and $a_J=0.042$ for the
state $\tilde\Phi_-$.
The main contribution to the integral in Eq.~(\ref{firstnegative})
comes from the vicinity of the tuning point $r_t$.
Unless $\delta$ and $\varepsilon_{\nu}$ are very large, the
wavefunction $\psi_g$ can be put equal to unity at distances $r\sim
r_t$.
Calculating the integral by using a linear approximation for the
potentials $V_{-}(r)$ and $V_{*-}(r)$ in the vicinity of $r_t$
we find
\begin{equation}
\nu_{pa}=\pi b_J\left(\frac{\Omega}{\delta}\right)^2
\frac{\Gamma^2}{(\delta-\varepsilon_\nu)^2+\Gamma^2}\Delta\varepsilon_{\nu}
\lambdabar^3 n_c n_s,  \label{firstnegative}
\end{equation}
For transitions to the state $\tilde\Phi_{JM}$ we have $b_J=a_J$, and for
transitions to the state $\tilde\Phi_-$ the coefficient $b_J=0.015$.

At resonance the rate of photoassociation Eq.~(\ref{firstnegative})
is larger by a factor $\sim \Delta\varepsilon_{\nu}/\Gamma$ than the
rate of radiative collisions given by Eq.~(\ref{rate2})
for similar but positive
detuning.
However, for large detuning the level spacing becomes very much larger
than $\Gamma$ and for most values of $\delta$ one will miss the
vibrational resonances.
The photoassociation rate is the smallest when the frequency detuning
is just in between two resonances, i.e., is of order
$\Delta\varepsilon_{\nu}$.
In this case the two nearest resonances will give the dominant
contribution and, assuming $\delta-\varepsilon_{\nu} \sim
\Delta\varepsilon_{\nu}$, we have
\begin{equation}         \label{secondnegative}
\nu_{pa}\sim (\Omega/\delta)^2 (\Gamma^2/\Delta\varepsilon_{\nu})
n_c n_s \lambdabar^3,
\end{equation}
which is smaller by factor of order
$\Gamma/\Delta\varepsilon_{\nu}$ than the rate of radiative collisions for
similar but positive detuning.
Accordingly, we have $\eta \sim
(\Lambda_b/\lambdabar)^3(\Delta\varepsilon_{\nu}/\Gamma)$.
Hence, for the negative $\delta$ case $\eta$ can in principle be
increased to
a value above unity for higher bath "temperatures" than in the case of
positive $\delta$.

In practice however, negative detuning does not lead to sub-Doppler
cooling in a VSCPT scheme on a J=1 to J=1 transition.
Therefore if we only rely on Doppler cooling to cool the bath,
the increase in $\eta$
resulting from the reduced rate of radiative collisions is counteracted
by the decrease of $\Lambda_b$. It is not clear whether this problem
can be easily dealt with.

\section{Conclusions}

We have shown that the operating conditions for an atom laser based
on VSCPT of He$^*$ are strongly limited by the loss mechanism associated
with radiative collisions.
For positive detuning it is necessary to precool the gas very close to the
recoil energy.
For negative detuning the situation is
more favorable but the lack of sub-Doppler cooling in VSCPT schemes
for negative $\delta$ may offset this advantage. We have shown that it
is not fundamentally impossible to realize the operating conditions
for an atom laser using VSCPT but in practice it may be rather difficult.
Clearly all other loss mechanisms should be carefully eliminated.

Although we did not analyse in detail other atom laser schemes, we
believe that in general it is crucial to take the effect of radiative
collisions into account when considering these models.

\section*{Acknowledgments}

We acknowledge fruitful discussions with M.W. Reynolds, Y.
Castin, J. Dalibard, and R.J.C. Spreeuw. This work was
supported by the Stichting voor Fundamenteel Onderzoek der Materie
(FOM), by the Nederlandse Organisatie voor Wetenschappelijk Onderzoek
(NWO) through project
NWO-047-003.036, by INTAS, and by the Russian Foundation for Basic Studies.

\begin{figure}
	\caption{
	Schematic depiction of the atom laser model. The bath and the
	system states are denoted $b$ and $s$ respectively. The pumping
	rate from the bath states is $\Gamma_b$ and the repumping from the
	system back into the bath is denoted by $\Gamma_s$.
	We assume that $\Gamma_s$ vanishes for zero momentum.
	The wiggly
	line represents a spontaneously emitted photon. The excited
	state involved in the process is denoted as $e$ and its inverse
	lifetime is $\Gamma$.
	}
	\label{1}
\end{figure}

\begin{figure}
	\caption{
	Level scheme involved in VSCPT cooling of He$^*$. The lower
	manifold is $2^3S_1$, the upper is $2^3P_1$. The numbers are
	the non-zero Clebsch-Gordan coefficients.
	}
	\label{2}
\end{figure}

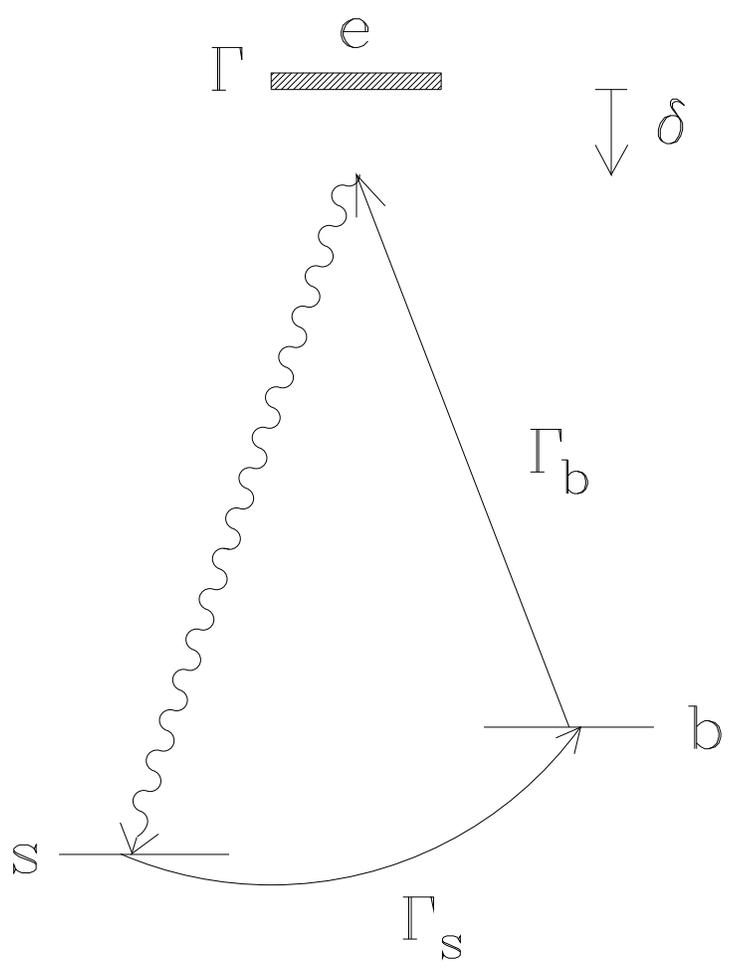
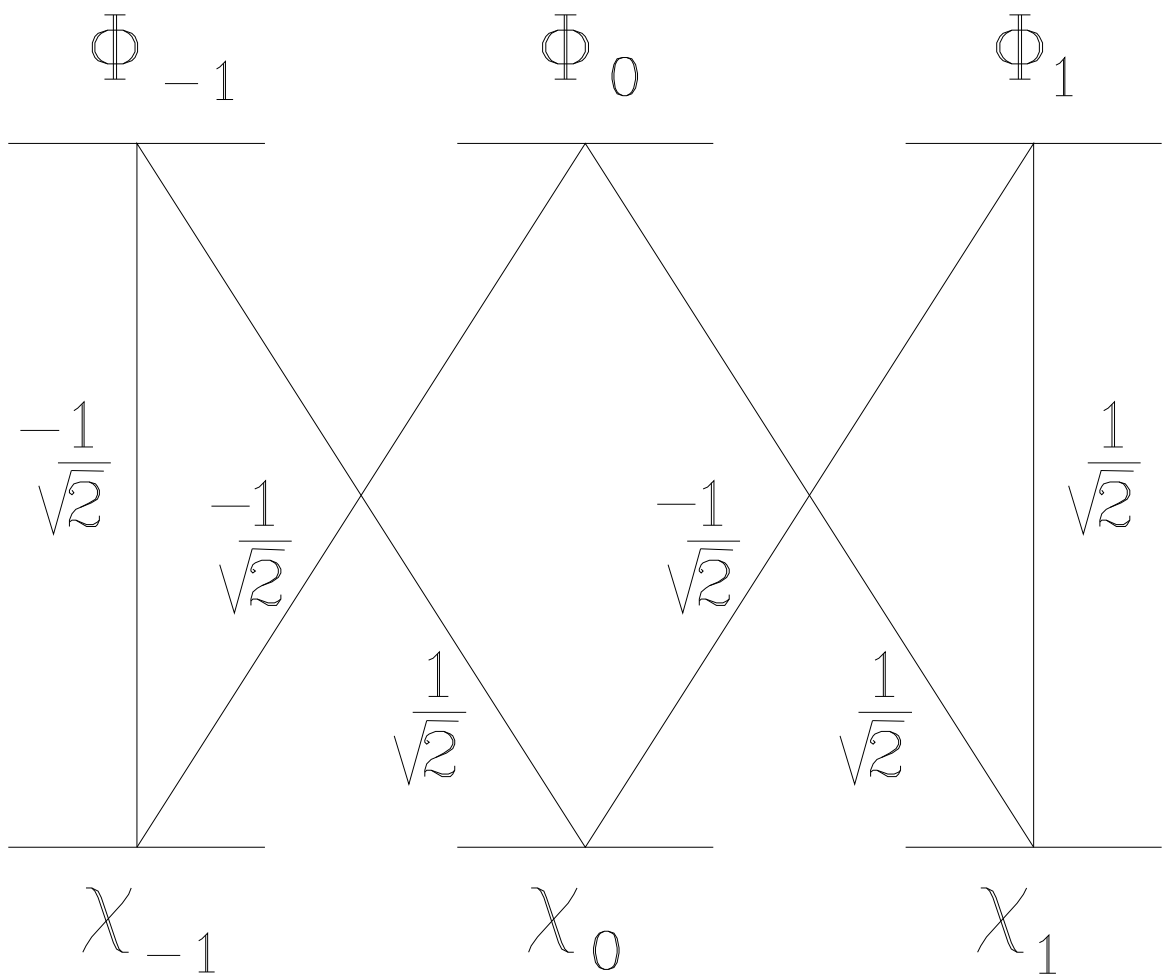
\end{document}